\documentclass[prl,twocolumn,floatfix,nobibnotes,superscriptaddress]{revtex4}
\usepackage{amsmath}
\usepackage{blindtext}
\usepackage{graphicx}
\usepackage[tight]{subfigure}
\usepackage[english]{babel}
\usepackage{csquotes}
\usepackage{amssymb}
\usepackage{amsmath}
\usepackage{graphicx}
\usepackage{subfigure}
\usepackage{gensymb}

\usepackage{color}
\usepackage[citecolor=blue,colorlinks=true]{hyper ref}

\begin{document}

\title{Spin-polaron formation and magnetic state diagram in La doped $CaMnO_3$}
\author{N. Bondarenko}
\affiliation{Division of Materials theory, Department of Physics and Astronomy, Uppsala University, Box 516, 75121 Uppsala, Sweden}
\author{Y. Kvashnin}
\affiliation{Division of Materials theory, Department of Physics and Astronomy, Uppsala University, Box 516, 75121 Uppsala, Sweden}
\author{J. Chico}
\affiliation{Division of Materials theory, Department of Physics and Astronomy, Uppsala University, Box 516, 75121 Uppsala, Sweden}
\author{A. Bergman}
\affiliation{Division of Materials theory, Department of Physics and Astronomy, Uppsala University, Box 516, 75121 Uppsala, Sweden}
\author{O. Eriksson}
\affiliation{Division of Materials theory, Department of Physics and Astronomy, Uppsala University, Box 516, 75121 Uppsala, Sweden}
\author{N.V. Skorodumova}
\affiliation{Division of Materials theory, Department of Physics and Astronomy, Uppsala University, Box 516, 75121 Uppsala, Sweden}
\affiliation{Multiscale Materials Modelling, Department of Materials Science and Engineering,Royal Institute of Technology, SE-100 44 Stockholm, Sweden}

\begin{abstract}

$La_xCa_{1-x}MnO_3$ (LCMO) has been studied in the framework of density functional theory (DFT) using Hubbard-U correction. We show that the formation of spin-polarons of different configurations is possible in the G-type antiferromagentic phase. We also show that the spin-polaron (SP) solutions are stabilized  due to an interplay of magnetic and lattice effects at lower La concentrations and mostly due to the lattice contribution at larger concentrations. Our results indicate that the development of SPs is unfavorable in the C- and A-type antiferromagnetic phases. The theoretically obtained magnetic state diagram is in good agreement with previously reported experimental results.

\end{abstract}
\pacs{later}

\maketitle

Perovskite $CaMnO_3$-$LaMnO_3$ (CMO-LMO) system exhibits an outstandingly rich magnetic and structural polymorphism \cite{Moreo}. $CaMnO_3$ (CMO) is an orthorhombic (Pnma) semiconductor with the band gap of 3.07 eV \cite{Jung}. Its magnetic ground state is the G-type antiferromagnetic (G-AFM) structure, where each spin-up (down) atom is surrounded by 6 spin-down (up) atoms. Such a magnetic ordering is thought to be governed by the super-exchange interaction along the $Mn^{4+}(t_{2g}^3) \uparrow-O(p)-Mn^{4+} (t_{2g}^3) \downarrow$ bond chains \cite{Kramers}. When trivalent $La^{3+}$ substitute atoms in the $Ca^{2+}$ sublattice extra valence electrons are added to the system. This extra charge can be redistributed among a large number of atoms or fully (or partially) localized at the d-orbitals of particular Mn atoms driving the double-exchange interaction in the mixed-valence $Mn^{3+}(e_{g}^1)\uparrow-O(p)-Mn^{4+}(t_{2g}^3) \uparrow$ bond alignment \cite{Zener}. The Hund coupling may then assist the spin flip at the central site of the magnetic octahedron  \cite{Herring}, thus forming a ferromagnetic (FM) 7-site droplet or the so-called 7-site spin-polaron (SP). Such 7-site SPs can be joined together in different configurations forming larger FM droplets, for example, involving 12-, 17- or 21-sites \cite{Meskine1,Chen}. Unlike classical polarons, where an electron is trapped due to a strong electron-lattice interaction \cite{Bondarenko1,Arapan},  spin polarons are thought to localize largely due to magnetic interaction \cite{Yakovlev,Emin}. However, cooperative spin-charge-lattice effects are also important for SPs as the formation of the $Mn^{3+}(e_{g}^1)$ state leads to the symmetry breaking by Jan-Teller distortions becoming more pronounced as the number of $Mn^{3+}$ atom increases. At a critical concentration the accumulated lattice deformation energy drives the magnetic transition to the C-type antiferromagnetic (C-AFM) state, which in $La$-doped $CaMnO_3$ is accompanied by the structural transition from Pnma orthorhombic to the $P_1/m$ monoclinic structure \cite{Moreo,Ling1,Meskine1,Meskine2,Santhosh,Loshkareva}.

\begin{figure*}[htb]
\begin{center}
\includegraphics[scale=0.85]{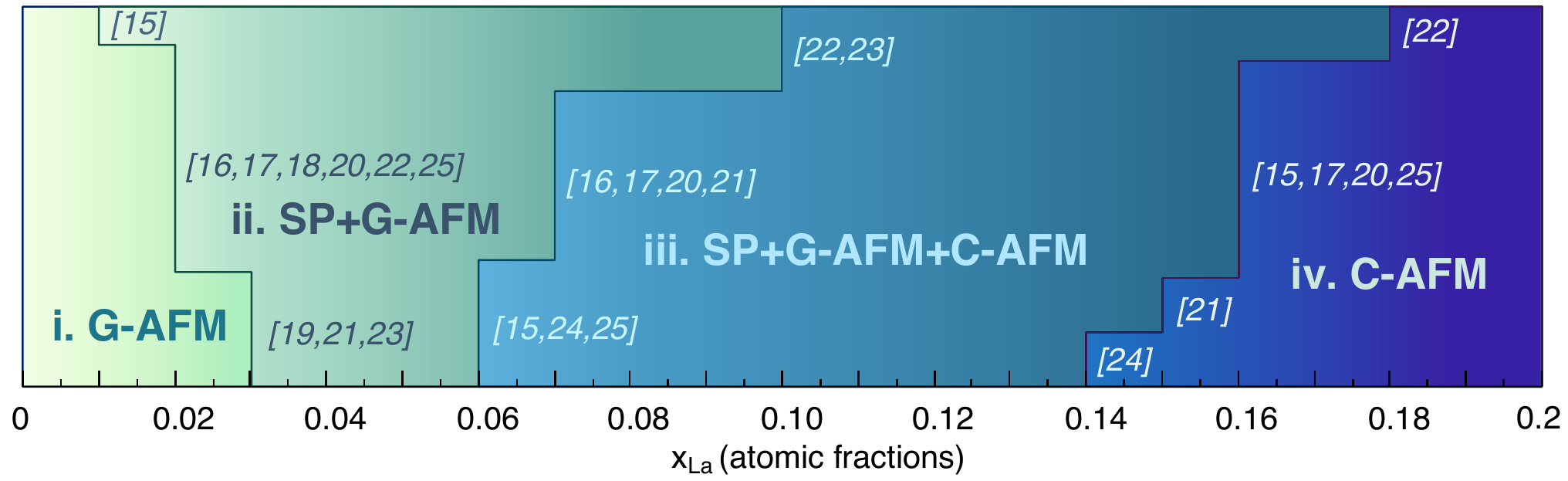}
\end{center}
\caption{\label{structure}(Color online) LCMO magnetic state diagram in $0<x_{La}<0.20$ range according to the following experimental methods: neutron powder diffraction \cite{Ling1}, magnetic properties \cite{Neumeier,Ling1,Chiorescu,Wang,Cortes}, electrical conductivity \cite{Lan}, Raman-scattering\cite{Granado1}, thermal conductivity \cite{Cohn}, electrical resistivity \cite{Cortes}, specific heat capacity \cite{Cornelius}, neutron scattering \cite{Ling2,Granado2}. The phase diagram is divided into four segments: i. G-AFM, ii. FM-droplets+G-AFM, iii. FM-droplets+G-AFM+C-AFM, iv. C-AFM. See text for details.} 
\label{fig5}
\end{figure*}

In Fig. 1 we summarize the available experimental data on the stability of the magnetic phases of $La_xCa_{1-x}MnO_3$ for $x_{La}<0.2$. The concentrations at which the magnetic transitions are reported to take place, vary depending on the experimental setups and applied methods, nonetheless, all the experiments clearly demonstrate the existence of four distinct regions (i-iv), described below.

i. Concentration range $0<x_{La}<0.01-0.03$. For these small concentrations the G-AFM magnetic structure of CMO is preserved but the physical properties of the oxide are noticeably affected by doping. In particular,  already La addition of 0.05-1 \% is enough to have a crucial impact on the Hall 
coefficient suggesting an increased mobility of charge carriers \cite{Chiorescu}. The measurements of electric conductivity confirm this suggestion reporting substantially higher values for the doped oxide compared to those for CMO \cite{Neumeier,Lan}. The magnetic saturation curve, $M_s(x_{La})$, measured for LCMO with $x_{La}<0.02 - 0.03 $, shows a slope of only 1 $\mu B/Mn $, which is much smaller than should be expected in the case of a FM droplet forming in the G-AFM matrix \cite{Neumeier, Cohn}. This finding speaks in favor of a mean field-like distribution of extra charge due to La-doping rather than the electron localization scenario.  

ii. Concentration range $0.01-0.03<x_{La}<0.06-0.10$. The oxide matrix still preserves the G-AFM order while a SP signature is also observed. The SP appearance has been detected by neutron powder diffraction, magnetization, Raman spectra and the heat conductivity  measurements \cite{Ling1,Neumeier,Cohn,Cortes, Wang,Cornelius,Granado1}. In contrast to the measurements done for smaller concentrations (region i), the slope of the $M_s(x_{La})$ curve now increases to  8 $\mu B/Mn $ \cite{Neumeier, Cohn}, confirming SP presence. Interestingly, the measured spontaneous magnetization is significantly smaller than a value to be expected in the case of full electron localization that suggests a partial character of localization \cite{Granado2,Wang, Loshkareva}. The neutron scattering measurements have distinguished isolated SPs of about 10.4 \AA~ in size separated by 41 \AA~ for $0.02 < x_{La} < 0.05$ and SPs up to 10.8 \AA~ in size separated by 24 \AA~for $0.05 < x_{La} < 0.10$  \cite{Granado2,Ling2}.  

iii. Concentration range $0.06-0.1<x_{La}<0.14-0.18$. Here one finds a complex mixture of magnetic structures including G-AFM with SPs and emerging C-AFM ordering, whose appearance is accompanied by the structural transition to the monoclinic ($P_1/m$) phase \cite{Cornelius,Wang,Ling1,Neumeier,Cohn,Cortes,Granado1,Granado2,Loshkareva}. 

iv. Concentration range $0.14-0.18<x_{La}<0.2$. According to numerous experiments \cite{Ling1,Wang,Neumeier,Cohn,Ling2,Cortes,Granado2} the C-AFM monoclinic phase is the only observed phase in this concentration interval.

The pioneer ab-initio calculation studying spin-polaron formation in  $La$-doped CMO has shown that  
the charge localization at the SP sites has $e_g$ character \cite{Meskine2}. The study of the electron doped CMO, done employing model Hamiltonians, has demonstrated the stability of the 7-site SP solution \cite{Meskine1,Meskine2,Chen}. The study of the model has also shown that beyond $x_{La}=0.045$, 7-site SP becomes unstable with respect to a FM spin order \cite{Chen}.

Here we present a detailed ab-initio description of the magnetic state diagram of $La_xCa_{1-x}MnO_3$ in the low-La concentration range. We consider the possibility for the SPs of different configurations to form in different AFM phases and analyze the role of magnetic and lattice contributions in their stabilization. 

In our study we used the DFT+U approach, employing the projector augmented wave  method \cite{Blochl} and the Perdew, Burke, Ernzerhof parametrization \cite{Perdew} of the exchange-correlation interaction as implemented in VASP \cite{Kresse}. The cutoff energy was 550 eV. The calculations were done for the 3x2x3 supercell containing 72 Ca, 72 Mn and  216 O atoms. The La concentration was varied in the range of $0.013<x_{La}<0.133$  by replacing different number of Ca atoms by La (1-10 La/supercell). A 2x2x2 Monkhorst-Pack k-point mesh, which resulted in 8 irreducible k-points, was used for the integration over the Brillouin zone. In order to find the equilibrium orthorhombic ratio a structural optimization was performed at each $x_{La}$ in the manner described in Ref.\cite{Bondarenko2}. The lattice optimization was done keeping the G-AFM magnetic order. SPs were formed in the magnetic lattice by  
flipping the spin on the central atom in one or several Mn octahedra and allowing only atomic positions to relax. 

The choice of the Hubbard U parameter is always an important issue in the calculations of complex oxides. Here we estimated the effective U parameter ($U_{eff}$=U-J \cite{Dudarev}) using the linear response method developed by Cococcioni \cite{Cococcioni}, which, depending on the choice of  the basis set, resulted in $U_{eff} (Mn_{3d})$ in the range of 3.45-4.23 eV. For our calculations, however, we utilized the  rotationally invariant approach \cite{Liechtenstein}. This approach was shown to be more appropriate for the description of complex magnetic structures \cite{Himmetoglu}. We used J=0.9 eV, most common value applied for this class of compounds \cite{Anisimov}. Further, we varied U in the range of U=0.9-8.9 eV that corresponds to U-J=0-8 eV. We found that $U<2.9$ eV  overestimate the stability of the G-AFM structure, while $U>4.9$ eV fail to describe the $Mn_{3d}-O_{2p}$ hybridization and  result in the stabilization of the long range FM structure. The U values between 2.9 and 4.9  eV reproduce qualitatively correct magnetic states, in agreement with experimental data \cite{Ling1,Granado2,Ling2}, (Fig.1). Based on the performed tests and the analysis of the parameters found in the literature \cite{Popovic,Hong,Aschaue}, we chose to use U=3.9 eV and J=0.9 eV in the calculations presented here. 

In Fig.2 we show the density of states (DOS) obtained for the 7- and 21-sites SPs formed in G-AFM matrix for two La concentrations. The excess electrons, donated by La, occupy a shoulder near the Fermi level, which is hardly visible for $x_{La}$=0.013 but becomes more evident as the La concentration increases (Fig.2). This shoulder consists of the $e_g$-states for all the studied SP configurations and La concentrations. The partial charge distribution shows that these $e_g$ states are mostly localized at the SP sites and have  $3z^2-r^2 , x^2-y^2$  character.  Somewhat larger degree of the $Mn({e^1_g})-O(p)$  hybridization is observed in the (101) plane as compared to the others. We have also performed hybrid functional calculations (HSE06) \cite{Heyd1,Hybrid} of the 7-site SP configuration and obtained very similar DOSs to the ones shown in Fig.2. 

\begin{figure}[htb]
\begin{center}
\includegraphics[scale=0.75]{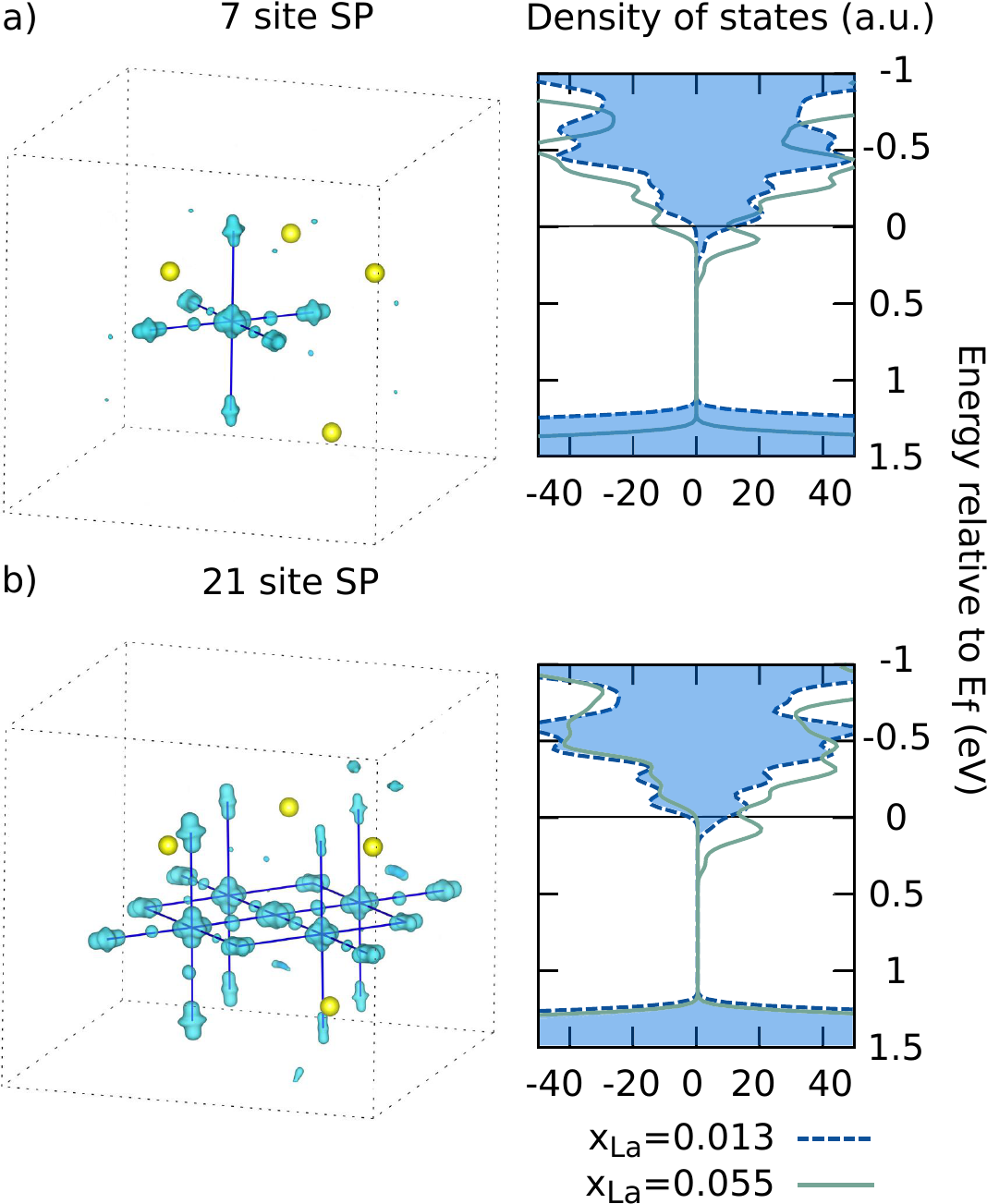}
\end{center}
\caption{\label{structure}(Color online) 
(a) Charge distribution of $e_g$ states (left panel) and the density of states (right panel) for 7-site SP for $x_{La}$=0.013 and $x_{La}$=0.055. 
(b) Charge distribution of $e_g$ states (left panel) and the density of states (right panel) for 21-site SP for $x_{La}$=0.013 and $x_{La}$=0.055. 
In the charge distributions only the states in the vicinity of the Fermi level are plotted. The La atoms are shown in yellow.  }
\label{fig2} 
\end{figure}

\begin{figure*}[htb]
\begin{center}
\includegraphics[scale=0.90]{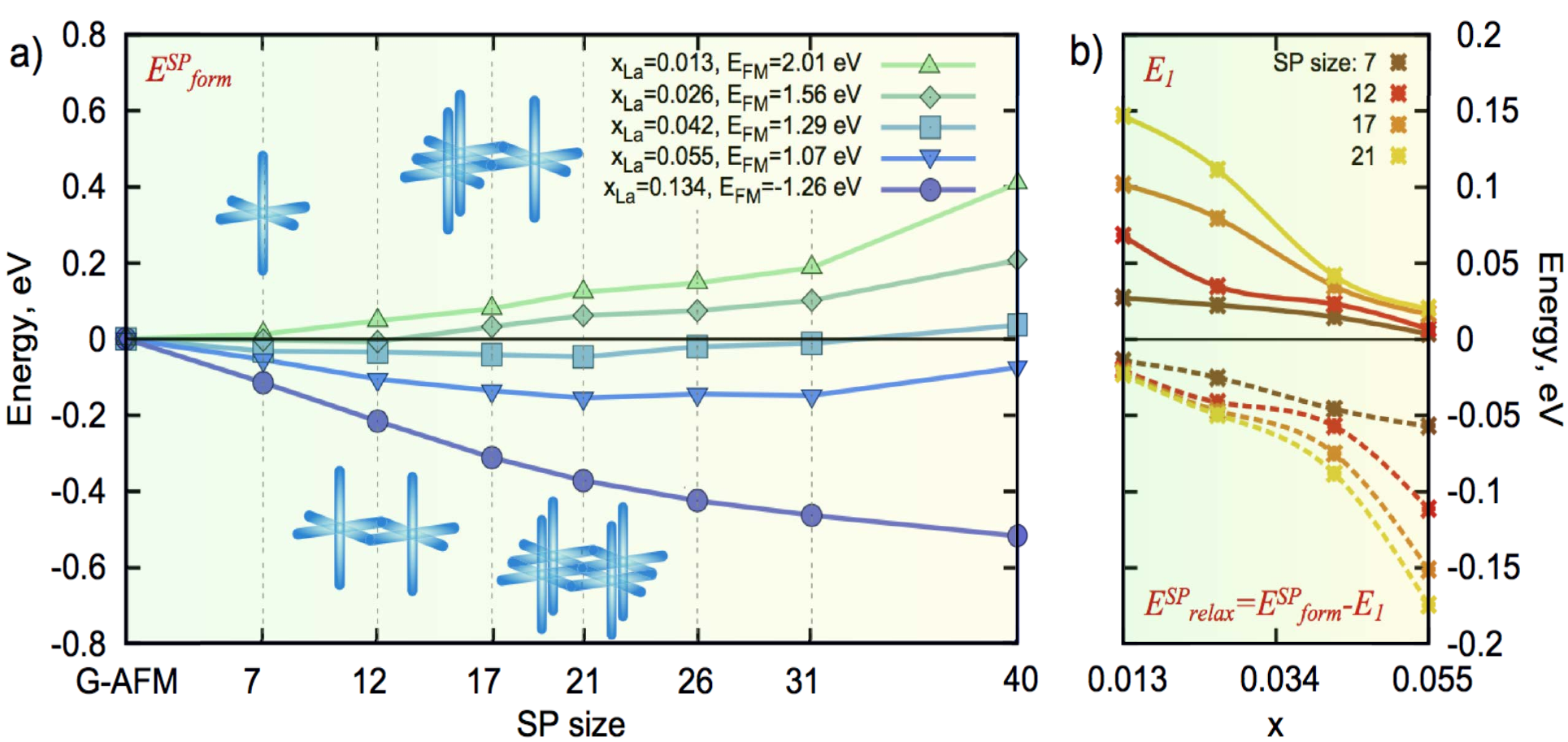}
\end{center}
\caption{\label{structure}(Color online) (a) Total energies of spin-polarons of various sizes formed in G-AFM  as a function of La concentration. The SP energies are given with respect to that of G-AFM ($E^{SP}_{form}$). For illustration, 7-, 12-, 17- and 21-site SPs are schematically shown. The energies of FM solutions are also listed for comparison. (b) Decomposition of $E^{SP}_{form}$ into the magnetic contribution ($E_1$, upper panel) and lattice relaxation contribution ($E^{SP}_{relax}$, lower panel). See text for details.}
\label{fig5} 
\end{figure*}

The formation energies of different SPs in the G-AFM phase calculated for five La concentrations are presented in Fig.3a. For $x_{La}=0.013$ the G-AFM solution is  more stable than the 7-site SP configuration by 13 meV. Larger SPs have higher energies and, therefore, even less stable with respect to G-AFM. However, already for $x_{La}=0.026$ the situation changes.  At this concentration the 7 and 12-site SPs become preferable by 3 meV and 6 meV, respectively. As La concentration increases  we observe further stabilization of spin polaron solutions. For $x_{La}=0.042$ and $x_{La}=0.055$ 21-site SP  has the lowest energy, lower than that of G-AFM  by 46 meV and 155 meV, respectively. This indicates a gradual development of a long range FM order, which we find to fully stabilize for the largest considered here concentration, $x_{La}=0.134 $ (Fig.3). Thus, we find the stabilization of the SP solutions between $x_{La}$=0.026 and 0.055. Most polaronic sites are situated in the (101) plane. The SPs prefer to form near the dopants. In particular, for $x_{La}$=0.013 we find that the energy of the 7-site SP increases by 18 meV as the SP-La separation distance changes from 3.34~\AA~to 6.56~\AA. 

Hypothetically, if one considers the case of a full localization of an extra electron at the SP sites one would expect to find additional 1~$\mu B$ per formed SP. The analysis of calculated Bader charges, however, shows that for small La concentrations the extra charge is practically totally smeared over the atoms of the oxide matrix. For $x_{La}$=0.013 (1 La/supercell), for example, only 0.06 $e^-$ for 7-site SP and 0.08 $e^-$ for 21-site SP of additional charge are found at the polaronic sites as compared to the rest of the sites in the supercell. In this case the SPs show an additional magnetic moment of 0.57~$ \mu_B/SP$ for 7-site SP and 0.47~$ \mu_B/SP$ for 21-site SP. For $x_{La}$=0.055 (4 La/supercell) we find extra 0.15 $e^-$ (7-site SP) and 0.20 $e^-$ (21-site SP) with the additional magnetic moments of 1.49 $ \mu_B/SP$ (7-site SP) and 1.28 $ \mu_B/SP$ (21-site SP). Therefore, we observe only partial charge localization at polaronic sites that agrees well with experimental findings \cite{Wang, Loshkareva, Granado2}.

To better understand the mechanism of the SP formation we try to separate lattice and magnetic contributions to the formation energy of the SP, $E^{SP}_{form}$, which is the difference between the total energies of the optimized supercell with SP and optimized G-AFM supercell. We estimate the contribution due to the spin subsystem reorganization by calculating the difference, $E_1$ (Fig.3b), between the energy of the supercell where all the atoms are frozen in the positions corresponding to the optimized G-AFM structure but the spins are arranged as in the SP, and the energy of the optimized G-AFM supercell. The difference between $E^{SP}_{form}$ and $E_1$ is the relaxation energy, $E^{SP}_{relax}$ (Fig.3b), or the energy contribution due to structural rearrangements around the SP. The corresponding energies are shown in Fig.3b for four SP sizes. The contributions vary with the SP size being much larger for 17- and 21-site configurations than for 7-site and 12-site SPs. Fig. 3b demonstrates that the lattice relaxation contribution for lower concentrations is rather small but it becomes substantially larger as La concentration increases. The magnetic contribution, on the contrary, decreases as more La is added. Therefore, the stabilization of smaller SPs (7- and 12-sites) is due to relatively small and comparable in value magnetic and lattice contributions. The stabilization of bigger SPs, however, is largely determined by the lattice contribution.

This conclusion is supported by the analysis of local lattice deformations. In particular, in the case of 7-site SP ($x_{La}$=0.013) the 6 Mn atoms surrounding the central Mn atom of SP shift away from it by $\sim 5~m$\AA~and the surrounding oxygens move away by $\sim 10~m$\AA~as compared to their positions in the G-AFM lattice. At the same time, Ca atoms come $\sim 5~m$\AA~closer to the SP center. For higher La concentrations ($x_{La}$=0.055), the displacements increase to $\sim 10~m$\AA~for Mn atoms, $\sim 50~m$\AA~for O atoms and $ \sim 20~m$\AA~for Ca atoms. At low concentrations mainly the atoms of the SP are displaced. For higher concentrations majority of the atoms of the supercell become displaced and the distortion amplitude increases.  

To clarify the total picture of the magnetic transitions in the low La concentration range we have examined four different magnetic phases experimentally observed in the CMO-LMO system: G-AFM, C-AFM, A-AFM and FM  \cite{Wollan}. For these calculations we have preserved the equilibrium orthorhombic lattice parameters but optimized the atomic positions for each magnetic  structure. The total energies of these phases with respect to the energy of the G-AFM phase are shown in Fig.4. In the low concentration range, $0<x_{La}<0.052$, the G-AFM state is most stable. In concentration interval $0.052<x_{La}<0.091$ we see the stabilization of the C-AFM phase. Finally, for $0.091<x_{La}<0.103$ the A-AFM phase emerges. If we now compare data in Fig.4 to those in Fig.1 we will discover that the theoretically predicted stability intervals of G-AFM and C-AFM are in good agreement with the experimental observations. The stabilization of the A-AFM structure in the considered concentration range, however, does not agree with experiment (Fig.1). 
As a matter of fact, this is the consequence of keeping the orthorhombic symmetry preserved.  Indeed, allowing the symmetry of the supercell to optimize at $x_{La}=0.11$ we find that it becomes monoclinic ($ \beta=0.91^{\circ}$,  a=14.89 \AA, b=14.86 \AA~and c= 15.12 \AA~).  The monoclinic C-AFM phase is  8 meV/f.u.  lower in energy than the A-AFM orthorhombic structure and 2.1 meV/f.u. lower than the A-AFM monoclinic structure. Therefore, our results confirm the stabilization of the monoclinic C-AFM phase in this concentration range.

\begin{figure}[htb]
\begin{center}
\includegraphics[scale=0.75]{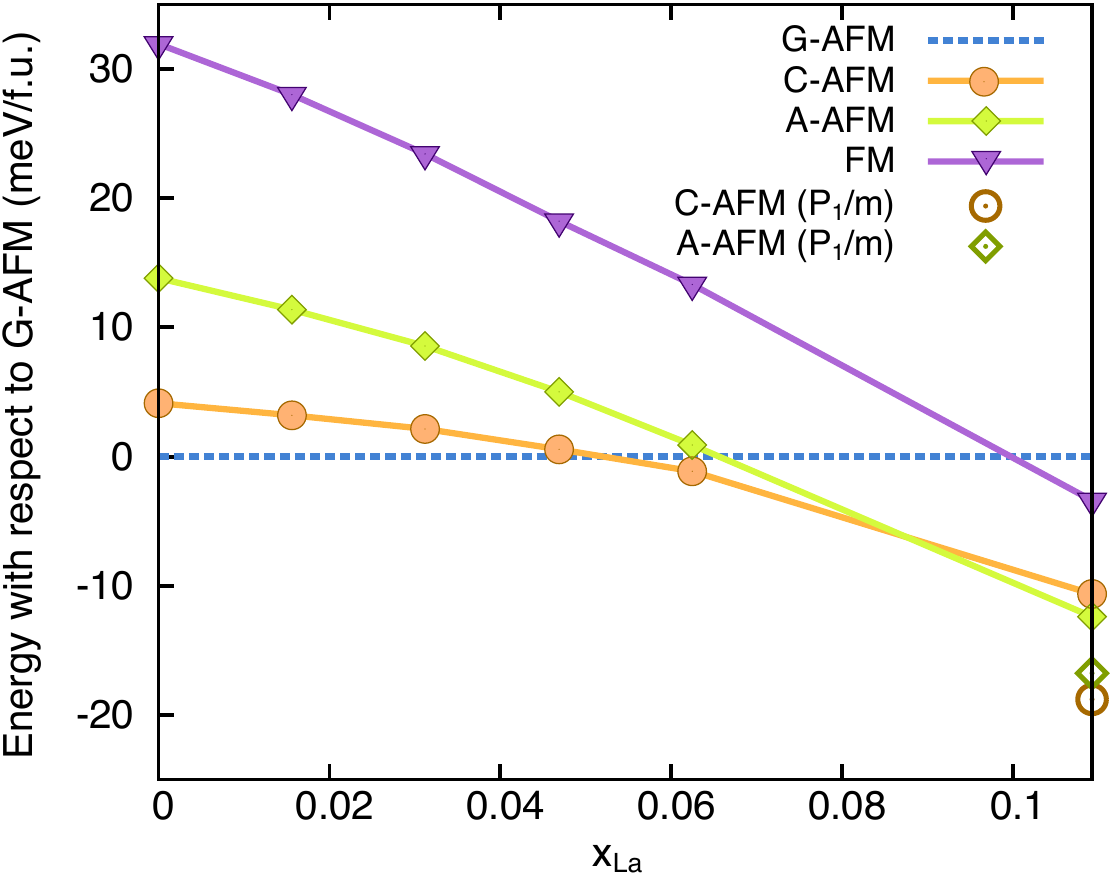}
\end{center}
\caption{\label{structure}(Color online) Energies of different magnetic phases as a function of $x_{La}$. We show total energies per formula unit with respect to that of G-AFM.}
\label{fig5} 
\end{figure}

Additionally,  we have checked the possibility for spin to flip and form some kind of local FM order in C-AFM and A-AFM phases. We notice, however, that only the G-AFM structure naturally supports the appearance of FM droplets assisted by the double-exchange  interaction along $Mn^{3+}(e_{g}^1)\uparrow-O(p)-Mn^{4+}(t_{2g}^3)\uparrow $ bonds, which can be achieved by the flip of a single spin. The arrangements of spins in the C- AFM and A-AFM phases  do not allow a local FM droplet to form by the same manipulation. Nevertheless we flipped one spin in each of the structures, relaxed the atomic positions and compared the resulting total energies to that of the corresponding AFM structure. We found that the energies of all these configurations are higher than those of their parental AFM structures by 20-60 meV depending on La concentration. 

To summarize, we have proposed an optimized approach, based on DFT+U, which allowed us to describe charge localization, spin polaron formation and magnetic polymorphism of LCMO. In particular, we have reported the reliable sizes of the SP as a function of La concentration, and demonstrated a  non-trivial relationship between the two. In addition, we have provided a microscopic understanding of the relative importance of the role of exchange and lattice effects for the formation of the spin-polarons in La doped $CaMnO_3$.

N.V.S. acknowledges the financial support of the Swedish Research Council (VR) (project 2014-5993). O.E. acknowledges support from the Swedish Research Council (VR) and the KAW foundation (grants 2013.0020 and 2012.0031). We thank the Swedish National Infrastructure for Computing (SNIC) for providing computational resources.


\begin{thebibliography}{}
\bibitem{Hybrid} The test caculations with hybrid functional were performed for a smaller unit cell (1 La, 15 Ca ,16 Mn and 48 O atoms).
\end{thebibliography}
\end{document}